\documentclass[conference]{IEEEtran}
\IEEEoverridecommandlockouts

\usepackage{eso-pic}

\usepackage{cite}
\usepackage{amsmath,amssymb,amsfonts}
\usepackage{algorithm}
\usepackage{algpseudocode}
\usepackage{graphicx}
\usepackage{textcomp}
\usepackage{listings}
\usepackage{xcolor}
\usepackage{booktabs}

\definecolor{jsonbackground}{rgb}{0.98,0.98,0.98}
\definecolor{jsonstring}{HTML}{1A1AA6}
\definecolor{jsonkey}{HTML}{0A5D32}
\definecolor{jsonnumber}{HTML}{A31515}
\definecolor{grayborder}{HTML}{E0E0E0}

\lstdefinelanguage{json}{
    basicstyle=\linespread{0.95}\ttfamily\footnotesize,
    backgroundcolor=\color{jsonbackground},
    frame=single,
    rulecolor=\color{grayborder},
    breaklines=true,
    showstringspaces=false,
    stringstyle=\color{jsonstring},
    keywordstyle=\color{jsonkey},
    commentstyle=\color{gray},
    numbers=none,
    morestring=[s]{"}{"},
    literate=
     *{0}{{{\color{jsonnumber}0}}}{1}
      {1}{{{\color{jsonnumber}1}}}{1}
      {2}{{{\color{jsonnumber}2}}}{1}
      {3}{{{\color{jsonnumber}3}}}{1}
      {4}{{{\color{jsonnumber}4}}}{1}
      {5}{{{\color{jsonnumber}5}}}{1}
      {6}{{{\color{jsonnumber}6}}}{1}
      {7}{{{\color{jsonnumber}7}}}{1}
      {8}{{{\color{jsonnumber}8}}}{1}
      {9}{{{\color{jsonnumber}9}}}{1}
}


\renewcommand\IEEEkeywordsname{Keywords}

\def\BibTeX{{\rm B\kern-.05em{\sc i\kern-.025em b}\kern-.08em
    T\kern-.1667em\lower.7ex\hbox{E}\kern-.125emX}}

\begin{document}
\AddToShipoutPictureBG*{%
  \AtPageUpperLeft{%
    \put(\LenToUnit{0.5\paperwidth},\LenToUnit{-1.2cm}){%
      \makebox[0pt][c]{%
        \begin{minipage}{0.9\paperwidth}
          \centering \footnotesize
          Hieu Duong, Eugene Levin, Todd Gary, and Long Nguyen. ``CaST: Causal Discovery via Spatio-Temporal Graphs in Disaster Tweets'', The 5th Workshop on Knowledge Graphs and Big Data, IEEE Big Data 2025 (KGBigdata 2025)
        \end{minipage}%
      }
    }
  }
  \AtPageLowerLeft{%
    \put(\LenToUnit{0.5\paperwidth},\LenToUnit{1.0cm}){%
      \makebox[0pt][c]{%
        \begin{minipage}{0.9\paperwidth}
          \centering \scriptsize
          \copyright 2025 IEEE. Personal use of this material is permitted. Permission from IEEE must be obtained for all other uses, in any current or future media, including reprinting/republishing this material for advertising or promotional purposes, creating new collective works, for resale or redistribution to servers or lists, or reuse of any copyrighted component of this work in other works.
        \end{minipage}%
      }
    }
  }
}

\title{CaST: Causal Discovery via Spatio-Temporal Graphs in Disaster Tweets}

\author{
\IEEEauthorblockN{Hieu Minh Duong\thanks{Corresponding author: hieu.duong@louisville.edu} and Long Nguyen}
\IEEEauthorblockA{\textit{J.B. Speed School of Engineering} \\
\textit{University of Louisville}\\
Louisville, KY, USA \\
\{hieu.duong, l.nguyen\}@louisville.edu}
\and
\IEEEauthorblockN{Eugene Levin and Todd Gary}
\IEEEauthorblockA{\textit{School of Applied Computational Sciences} \\
\textit{Meharry Medical College}\\
Nashville, TN, USA \\
\{elevin, tpgary\}@mmc.edu}
}

\maketitle

\begin{abstract}
Understanding causality between real-world events from social media is essential for situational awareness, yet existing causal discovery methods often overlook the interplay between semantic, spatial, and temporal contexts. We propose CaST: Causal Discovery via Spatio-Temporal Graphs, a unified framework for causal discovery in disaster domain that integrates semantic similarity and spatio-temporal proximity using Large Language Models (LLMs) pretrained on disaster datasets. CaST constructs an event graph for each window of tweets. Each event extracted from tweets is represented as a node embedding enriched with its contextual semantics, geographic coordinates, and temporal features. These event nodes are then connected to form a spatio-temporal event graph, which is processed using a multi-head Graph Attention Network (GAT) \cite{gat} to learn directed causal relationships. We construct an in-house dataset of approximately 167K disaster-related tweets collected during Hurricane Harvey and annotated following the MAVEN-ERE schema. Experimental results show that CaST achieves superior performance over both traditional and state-of-the-art methods. Ablation studies further confirm that incorporating spatial and temporal signals substantially improves both recall and stability during training. Overall, CaST demonstrates that integrating spatio-temporal reasoning into event graphs enables more robust and interpretable causal discovery in disaster-related social media text.
\end{abstract}


\section{Introduction}
\begin{figure}
    \centering
    \includegraphics[width=\linewidth]{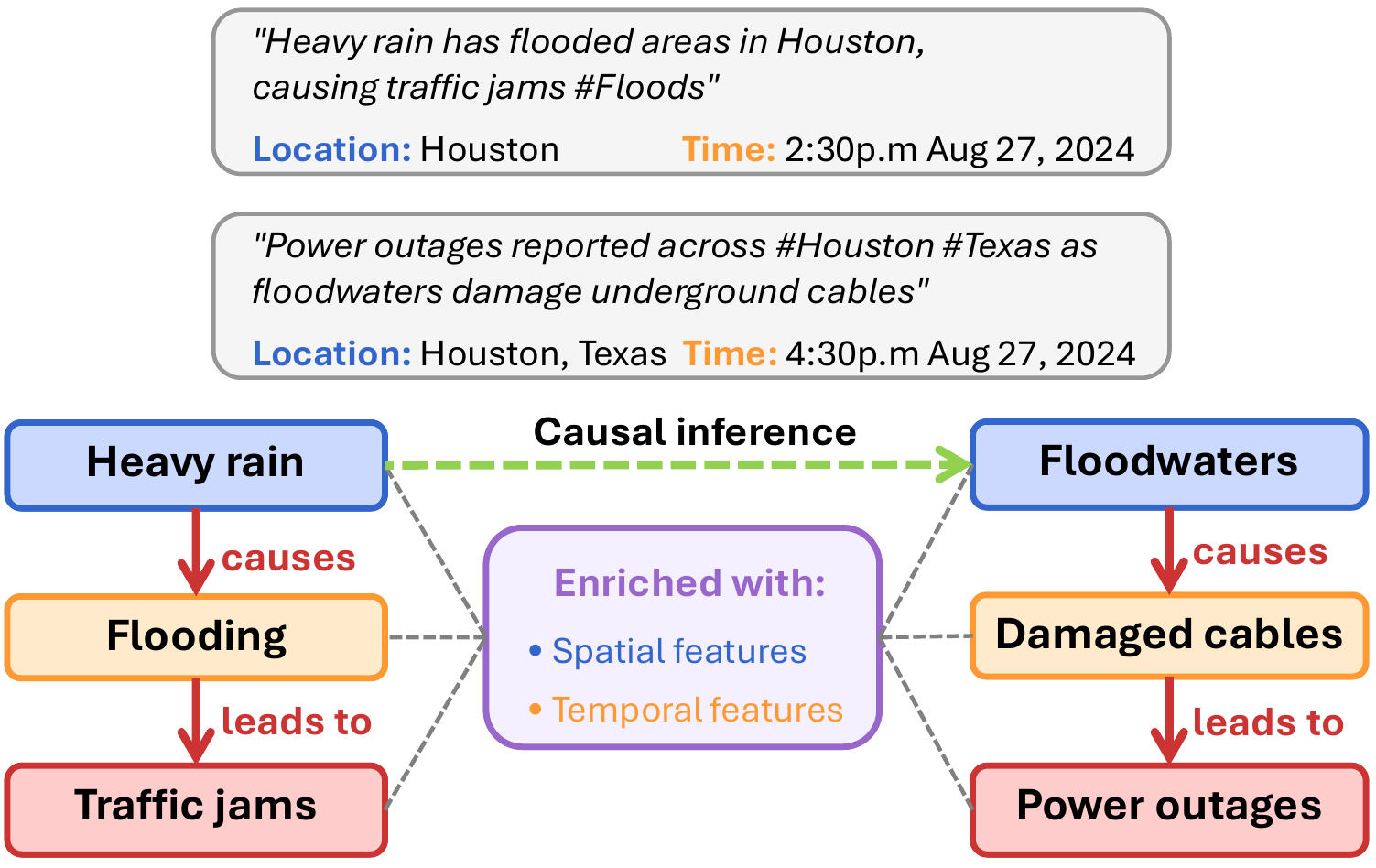}
    \caption{Example of causal relationships in disaster tweets with spatio-temporal context. Two tweets from the same location (Houston) and close temporal proximity (2 hours apart) contain causally related events. The green dashed arrow illustrates the causal inference between "Heavy rain" and "Floodwaters" across tweets, enabled by shared spatio-temporal features. Gray dashed lines show how spatial and temporal context enriches event representations to improve causal discovery.}
    \label{fig:example}
\end{figure}

Causal discovery aims to identify cause–effect relationships among events from observational data, offering deeper insights into how complex phenomena unfold. In disaster scenarios, understanding causality across space and time is critical for improving situational awareness, forecasting cascading impacts, and supporting decision-making for emergency response. Social media platforms such as Twitter have become valuable data sources in this context, as they provide real-time, fine-grained signals of unfolding disaster events, ranging from infrastructure damage to community-level responses. These data streams often reflect complex, interdependent phenomena that evolve rapidly across locations and timescales. Figure \ref{fig:example} illustrates how disaster-related events unfold across time and space in social media text. In this example, one tweet describes heavy rain leading to flooding and traffic jams, while another later tweet reports floodwaters causing damaged cables and power outages in the same region. Although these tweets are posted separately, they represent a continuous chain of disaster effects occurring within a short temporal window and close spatial proximity. Such examples highlight the need for spatio-temporal reasoning in causal discovery: existing text-based approaches often capture only intra-tweet relations, overlooking cross-tweet causal propagation that frequently occurs in real-world disaster scenarios.

However, inferring causality from such unstructured, dynamic, and context-rich text remains a challenging task. Traditional causal discovery approaches, including statistical and graph-based models, often fail to capture the intricate spatial and temporal dependencies that characterize real-world disaster information. Recent advancements in Large Language Models (LLMs) such as BERT and GPT have shown exceptional capabilities in semantic understanding, enabling extraction of event representations from unstructured text. Yet, most existing LLM-based methods for causal inference focus solely on textual relationships and ignore the spatio-temporal interdependencies among events, limiting their ability to capture the real-world propagation patterns typical of disasters. In particular, events such as \emph{floods causing power outages}, \emph{aftershocks triggering new collapses}, or \emph{road closures leading to rescue delays} require both temporal reasoning (when events happen) and spatial awareness (where they occur) to be causally understood. Integrating these dimensions is therefore essential to identify how disaster effects spread and interact.

Despite recent advances, current approaches often fall short in integrating both spatio-temporal reasoning within a unified causal discovery pipeline. Many graph-based models overlook temporal dependencies or treat cross-document causality as a secondary task, while prompt-based LLM methods, though flexible, lack structural consistency and are prone to hallucination. Likewise, methods that inject external knowledge or event semantics often do so statically, without modeling how events evolve and interact across time and space—an essential aspect in domains like disaster response, where causal chains unfold dynamically across locations and hours. In large-scale events such as hurricanes, floods, or wildfires, understanding where and when cascading effects occur is vital for situational awareness and coordinated response. However, few existing methods explicitly capture such spatio-temporal propagation patterns in social media streams. 
To address these challenges, we aim to develop a unified framework that integrates semantic, spatial, and temporal knowledge into a single graph-learning paradigm. In this framework, each event extracted from social media text is represented as a node enriched not only with linguistic features from LLM embeddings but also with contextual signals derived from its spatial location and temporal occurrence. The event graph is then constructed and expanded along two dimensions: (i) \textit{Spatial neighbors}, capturing proximity in location, and (ii) \textit{Temporal neighbors}, capturing alignment in time. This enriched graph is subsequently processed using a multi-head Graph Attention Network (GAT), which enables dynamic attention across spatial and temporal relations. Through this design, CaST bridges the gap between text-based causality and real-world disaster dynamics, allowing the model to learn context-aware causal representations that better reflect event evolution across time and space. Our contributions are summarized as follows:
\begin{itemize}
    \item We propose CaST (Causal Discovery via Spatio-Temporal Graphs), a novel framework for causal discovery that unifies spatial and temporal relationships into a spatio-temporal event graph derived from disaster-related social media data. Unlike prior methods that rely purely on textual cues, CaST explicitly captures how disaster-related events unfold across time and locations.
    \item We construct a large-scale dataset of approximately 167K disaster-related tweets, annotated following the MAVEN-ERE schema with extensions for temporal, spatial, and causal information. This dataset enables fine-grained evaluation of causal inference in real-world, noisy social media contexts.
    \item We assess CaST against eight strong baselines, ranging from traditional approaches to graph-based, neural, and prompt-based models. Results show that CaST achieves the best balance between precision and recall, outperforming prior methods in both causal link identification and generalization stability. Ablation studies further confirm the importance of spatial and temporal representations in enhancing causal inference.
\end{itemize}

\section{Related Work}
\subsection{Early Approaches for Causal Discovery}
Early methods identified causal relations in text using patterns and linguistic cues. Khoo et al.\ \cite{khoo1998} applied manual templates such as ``X leads to Y'' to newspaper articles to extract cause--effect pairs. Supervised feature-based classifiers were later introduced: Blanco et al.\ \cite{blanco2008} trained a classifier on lexical and syntactic features to detect causal relations. Researchers also began creating evaluation datasets; the SemEval-2010 Task 8 included Cause–Effect as one of several semantic relation classes. Girju \cite{girju2003} developed an early question–answering system for causal QA, using semantic parsing and WordNet knowledge to detect causality in text. With the rise of machine learning and larger datasets, causal extraction methods shifted toward neural networks and external knowledge integration. Cao et al.\ \cite{cao2021} proposed a knowledge-enriched model (LSIN) that incorporates external graphs for event causality identification. Their Latent Structure Induction Networks use two modules: a Descriptive Graph Induction, which pulls descriptive attributes of events from knowledge bases, and a Relational Graph Induction, which learns latent multi-hop connections between events. 

Data augmentation and transfer learning were also explored. Cao et al.\ and Liu et al.\ generated synthetic causal sentences or mined additional training examples from external corpora. Cao et al.\ \cite{cao2021} leveraged patterns from lexical knowledge bases (WordNet, VerbNet) to create paraphrases of causal statements and used contrastive learning to teach a model invariant features of causality. Neural models were further extended to jointly model related tasks. Shen et al.\ \cite{shen2022} introduced a prompt-based multi-task learning approach for event causality identification. They crafted auxiliary prompt tasks that a pretrained language model must answer about the presence of causal cues and plausible causal links, drawing on latent knowledge of causation in the language model. Zuo et al. \cite{zuo2021} present LearnDA, a dual-learning augmentation framework for causal sentence generation. Transfer-learning strategies are examined by Anuyah et al. \cite{anuyah2025}, who show that combining data from multiple domains improves causal relation extraction performance. More recently, Chun et al.\ \cite{chun2025} leverage pretrained language models (e.g., RoBERTa, T5) for data augmentation in event causality tasks, demonstrating benefits even under label imbalance. These studies highlight evolving trends: beyond vanilla supervised models, augmentation and cross-domain transfer hold promise for improving causality extraction in low-resource or implicit-causal settings.

\subsection{Graph-Based and Joint Reasoning Methods}

Recent state-of-the-art approaches emphasize structured reasoning using graphs to represent events and their relations. Instead of treating each candidate event pair independently, these methods model global structures where multiple causal links interact and support logical consistency across a document.
Gao et al.\ \cite{gao2019} introduced one of the earliest structured approaches, building an Integer Linear Programming (ILP) framework to enforce global constraints on causal relations at the document level. Their system identified key central events and applied constraints such as transitivity and consistency with discourse cues to capture coherent causal chains within long narratives.

Following this, neural graph-based models were developed to integrate structured reasoning directly into the model architecture. Graph Neural Networks (GNNs) became the primary tool for this purpose. Models such as those by Phu and Nguyen \cite{gcn} and Fan et al.\ \cite{fan2022} represented each document as an event graph, where nodes correspond to events and edges represent potential causal or semantic relations. Through message passing, information from surrounding events, coreference links, and discourse relations informed the prediction of causal links within and across sentences. Liu et al.\ \cite{identifying} extended this idea by constructing directed causal graphs and iteratively refining them through an iterative Learning and Identifying Framework (iLIF), which first identified high-confidence causal links and then updated event representations through a graph encoder before repeating the process.

Building on these developments, Pu et al.\ \cite{pu2024} proposed a joint framework to extract events and causal relations simultaneously. Their model employed a heterogeneous relation graph that included event–event edges for causality, as well as event–argument and argument–argument edges for shared entities and semantic roles. Event types and words formed the graph nodes, and a GNN captured the interactions among event identity, arguments, and candidate causal connections. They also introduced a multi-channel label enhancement strategy to represent the cause and effect roles within this unified architecture.

To enrich the semantic structure of graph-based models, Hu et al.\ \cite{hu2023} incorporated Abstract Meaning Representation (AMR) to provide deeper semantic context for causal inference. Sentences were converted into AMR graphs encoding the conceptual and relational structure of events, and information from AMR subgraphs connecting event mentions was aggregated to construct a semantic reasoning graph. Their model, SemSIn, used a GNN to integrate event-centric structures with an LSTM-based path encoder for event-associated structures, allowing the representation of implicit relations through underlying semantic links.

\subsection{Spatio-Temporal and Domain-Specific Causal Discovery}

A special subclass of causal text mining focuses on integrating temporal and spatial information to understand when and where events occur. Mirza and Tonelli were among the first to jointly analyze temporal and causal relations in text \cite{mirza2014}. They proposed a framework that links event causality with temporal ordering, using the principle that causes generally precede effects in time and employing temporal relations as features for causal inference.

Liu et al.\ (2023) \cite{ppat} introduced PPAT (Progressive Graph Pairwise Attention Network) , which refines event pair representations through progressive attention layers and graph-based message passing. PPAT models causal directionality more effectively than GCN-based methods by learning hierarchical attention from token-level to event-level features. Liu et al.\ (2025) \cite{liu2025} proposed a model that jointly learns event causality and temporality. Their framework uses a dual-channel neural network, where one channel identifies causal relations between events and the other detects temporal before/after relations. The model constructs an event causality graph in which nodes are weighted by temporal salience, allowing temporal signals to inform causal reasoning and providing a unified representation of event dependencies.

Spatio-temporal modeling has also been extended to domain-specific contexts such as disaster analysis and social media monitoring. Dong et al.\ \cite{dong2025} developed a real-time framework for extracting cascading disaster effects from Twitter streams. Their system continuously collects posts during disasters, applies co-word analysis to detect emerging topics, identifies potential causal chains, and enriches them by extracting geolocation references to visualize geographically distributed consequences. In another domain, Lenti et al.\ \cite{lenti2025} investigated causal discovery in social movements related to climate activism. They constructed a probabilistic causal graph integrating Reddit discussion data with external event information to model the factors influencing protest participation. The framework captures multi-factor causal influences, representing relationships such as interactions with activists, media exposure, and engagement intensity.

\subsection{Large Language Models for Causal Discovery}

Large Language Models (LLMs) such as GPT-3.5 and GPT-4 have introduced a new direction for causal reasoning from text. Kiciman et al. \cite{kiciman2024} evaluated these models across several benchmarks, including pairwise causal discovery, counterfactual reasoning, and event causality tasks. Their study demonstrated that LLMs can infer causal relationships directly from textual descriptions, generate causal graphs, and provide natural language explanations of causal links, reflecting their internalized commonsense knowledge acquired through large-scale pretraining.

Subsequent research has explored combining LLMs with traditional causal inference algorithms. Long et al.\ (2023) demonstrate this by using GPT-generated causal statements as priors or constraints for Bayesian network learners as in NLP applications, LLMs have been employed to extract candidate causal graphs from corpora that can later be refined through statistical modeling or human feedback \cite{long2023}. This hybrid approach leverages the linguistic and world knowledge capabilities of LLMs while connecting them with structured causal frameworks. Recent work has further extended this paradigm by integrating structured representations directly into LLM-based reasoning. Shyalika et al.\ \cite{shyalika2024} incorporated Causal Event Graphs as contextual input to LLMs to enhance spatial–temporal event reasoning. In this setup, LLMs process both textual and graph-based representations, combining natural language understanding with explicit relational structures for more interpretable causal inference.

\section{Methodology}
\subsection{Problem Formulation}
According to Hume’s definition of causality \cite{Holland01121986}, three fundamental conditions must hold: A causes B if (1) A precedes B in time, (2) A and B are contiguous in space and time, and (3) A and B co-occur or neither occurs.
Building upon this philosophical grounding, our framework operationalizes causality through explicit temporal precedence, spatial proximity, and contextual co-occurrence among detected events in disaster-related tweets. We formalize causal discovery from disaster tweets as an edge classification problem on a spatio-temporal event graph $\mathcal{G}=(\mathcal{V},\mathcal{E})$, where nodes represent extracted events enriched with semantic, spatial, and temporal features, and edges denote candidate causal connections. The task is to predict whether a directed edge $e_i \rightarrow e_j$ exists between two events based on their contextual and structural representations. 

\subsection{Framework Overview}
\begin{figure*}[t]
    \centering
    \includegraphics[width=\textwidth]{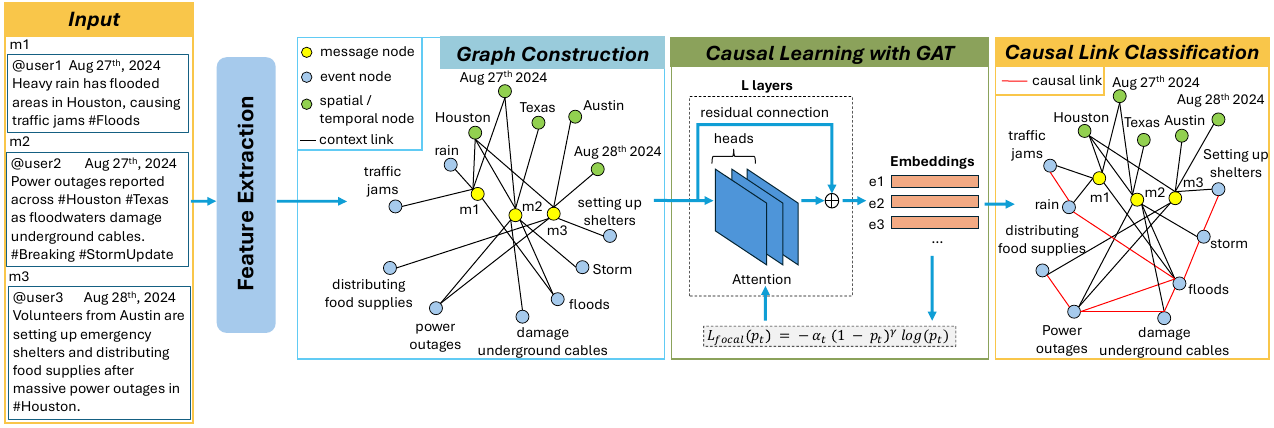}
    \caption{Overview of the proposed CaST framework for causal discovery in disaster tweets. 
    The Feature Extraction module extracts events along with triggers and contextual attributes (spatial and temporal attributes), encodes them as heterogeneous nodes. 
    A spatio-temporal graph is then constructed where black edges denote contextual link (semantic, spatial, or temporal relation), and red edges indicate predicted causal links in the output. 
    The Graph Attention Network propagates information across these edges to learn context-aware representations that enable accurate causal link classification.}
    \label{fig:framework}
\end{figure*}

Figure \ref{fig:framework} illustrates the overall architecture of the proposed CaST framework. The procedural steps are organized into three modules—Feature Extraction (Algorithm \ref{alg:feature}), Spatio-Temporal Graph Construction (Algorithm \ref{alg:graph}), and Causal Learning with GAT (Algorithm \ref{alg:gat}). CaST begins by extracting event triggers, arguments, and their spatial and temporal attributes from disaster-related tweets. These elements are encoded into heterogeneous node representations and serve as the basis for constructing a unified spatio-temporal event graph. The graph incorporates semantic, spatial, and temporal relationships that link events across tweets within each time window. Once the graph is built, the model applies a multi-head Graph Attention Network (GAT), which enables dynamic attention across spatial and temporal relations. A final classifier then predicts directional causal links between event pairs, allowing CaST to infer causal dynamics that are grounded in both linguistic content and spatio-temporal context.

Each tweet is processed to extract event triggers and their arguments, along with associated spatial and temporal cues. These elements form heterogeneous nodes: event nodes capture semantic content, while spatial and temporal nodes encode contextual information such as location and timestamp. Before graph learning, CaST combines these features into unified event representations that integrate semantic, spatial, and temporal signals. The spatio-temporal event graph contains nodes for events and contextual attributes, with edges representing semantic, spatial, and temporal relations. The Graph Attention Network then propagates and aligns information across these connections, enabling the model to learn dependencies that reflect how disaster-related events unfold. Finally, a causal link classifier determines whether a directed causal relation exists between any event pair, producing a structured representation of causal chains across time and space.

\subsection{Feature Extraction}
Algorithm \ref{alg:feature} outlines the process used to extract event triggers, arguments, and their associated spatial–temporal attributes from the raw tweet stream. The steps in Algorithm~\ref{alg:feature} are implemented as follows.
\begin{algorithm}[!htbp]
\caption{Feature Extraction}
\label{alg:feature}
\begin{algorithmic}[1]
\Require Tweet set $\mathcal{T}$
\Ensure Extracted events and initial event features

\For{each tweet $t_i \in \mathcal{T}$}
    \State Extract event triggers and arguments using NER and patterns.
    \State Obtain temporal features $\tau_i$ and spatial features $s_i$.
    \State Construct initial event representations from semantic, spatial, and temporal features.
\EndFor

\State \Return extracted events and event representations
\end{algorithmic}
\end{algorithm}

Given a tweet $t_i$ containing a sequence of tokens $\{w_1, w_2, \dots, w_n\}$, each annotated event trigger $e_j$ is encoded into a dense representation using the CrisisTransformer model \cite{crisis}.  
The contextualized embedding $\mathbf{v}_{e_j}$ is derived by combining the trigger token embedding and the [CLS] token embedding as:
\begin{equation}
    \mathbf{v}_{e_j} = \alpha \cdot \mathbf{v}_{\text{trigger}} + (1 - \alpha) \cdot \mathbf{v}_{\text{[CLS]}},
\end{equation}
where $\alpha = 0.7$ is empirically selected to balance local and global context. Each tweet is also associated with metadata providing temporal and spatial cues. 

Temporal features $\mathbf{t}_i$ are normalized indicators derived from the posting timestamp:
\begin{equation}
    \mathbf{t}_i = \left[ \frac{h}{24}, \frac{d}{7}, \frac{m}{12}, \frac{day}{31} \right],
\end{equation}
where $h$, $d$, $m$, and $day$ represent the hour, weekday, month, and day components, respectively.  
Spatial features $\mathbf{s}_i$ capture both coordinate-based and text-based location signals extracted from each tweet:
\begin{equation}
    \mathbf{s}_i = [g_1, g_2, l_1, l_2, \hat{\phi}, \hat{\lambda}],
\end{equation}
where $g_1$ and $g_2$ are binary indicators denoting the presence of geolocation metadata (from tweet coordinates or bounding boxes) and explicit textual location mentions, respectively.  
$l_1$ represents the normalized count of distinct location mentions extracted from the tweet text, while $(\hat{\phi}, \hat{\lambda})$ denote the latitude and longitude values normalized to $[-1, 1]$ by dividing by $90$ and $180$, respectively.  
When geolocation information is missing, coordinate values are set to zero.  
This formulation enables CaST to incorporate both explicit spatial references in text and implicit geographic metadata where available.

The extracted event, spatial, and temporal embeddings correspond to distinct node types that form the heterogeneous graph.  
Before input to the GAT model, these representations are projected into a shared latent space and concatenated to form unified node features:
\begin{equation}
    \mathbf{x}_{e_j} = [\mathbf{v}_{e_j} \, \Vert \, \mathbf{t}_i \, \Vert \, \mathbf{s}_i],
\end{equation}
where $\mathbf{v}_{e_j}$, $\mathbf{t}_i$, and $\mathbf{s}_i$ denote the event, temporal, and spatial feature vectors, respectively.  
This fusion ensures that each event node encodes both semantic and contextual information, enabling the GAT to learn richer spatio-temporal dependencies in subsequent causal graph learning.

\subsection{Graph Construction}
Algorithm \ref{alg:graph} details how these extracted elements are organized into a unified spatio-temporal event graph within each temporal window.

\begin{algorithm}[!htbp]
\caption{Spatio-Temporal Graph Construction}
\label{alg:graph}
\begin{algorithmic}[1]
\Require Processed tweets $\mathcal{T}$, time window $w_t$
\Ensure Set of spatio-temporal graphs $\{\mathcal{G}\}$

\State Partition tweets into chronological windows of size $w_t$.
\For{each window}
    \State Initialize graph $\mathcal{G}$.
    \For{each tweet $t_i$ in the window}
        \State Add event nodes to $\mathcal{G}$.
        \State Link each event to its spatial and temporal nodes.
        \State Add edges between events co-occurring in the same tweet.
    \EndFor
    \State Add $\mathcal{G}$ to the graph set.
\EndFor

\State \Return $\{\mathcal{G}\}$
\end{algorithmic}
\end{algorithm}

The Graph Construction module transforms event representations into a unified spatio-temporal graph that models relationships among events, locations, and time references.
Each tweet contributes: (1) \textit{event nodes} representing detected event triggers and their arguments, (2) \textit{spatial nodes} corresponding to extracted geographic entities such as cities or regions, and (3) \textit{temporal nodes} denoting timestamps derived from tweet metadata or explicit temporal expressions.
These nodes collectively encode the semantic, spatial, and temporal characteristics of disaster-related discourse. 

To preserve chronological dependencies and manage scalability, CaST constructs individual graphs within a fixed temporal window of tweets (e.g., several hours). Tweets falling within the same window are grouped to form one graph, ensuring that events close in time are contextually connected while maintaining temporal ordering across graphs. This sliding-window approach captures local causal dynamics and supports incremental learning as new tweets arrive, reflecting the evolving nature of disaster events.

Edges are established to represent interdependencies among nodes.
Contextual links connect event nodes that co-occur within the same tweet or exhibit high semantic similarity.
Spatial links connect events occurring in geographically proximate regions, determined via geolocation metadata or named-entity alignment.
Temporal links connect events according to their chronological order within each graph window, ensuring that causal direction respects temporal precedence.
This design operationalizes Hume’s classical notions of causation—temporal priority and spatial contiguity—within a data-driven framework. 

The resulting graph $\mathcal{G} = (\mathcal{V}, \mathcal{E})$ is heterogeneous, where $\mathcal{V}$ includes event, spatial, and temporal nodes, and $\mathcal{E}$ comprises semantic, spatial, and temporal relations.
Each node $v_i \in \mathcal{V}$ is initialized with its corresponding embedding vector (semantic from the language model, spatial from location encoding, and temporal from timestamp encoding).
These heterogeneous embeddings are aligned into a shared feature space and subsequently passed to the Graph Attention Network for relational reasoning and causal inference. This window-based graph construction allows CaST to model both local and evolving causal interactions among disaster events, effectively balancing temporal continuity and computational efficiency.

\subsection{Causal Learning with Multi-head GAT model}
Algorithm~\ref{alg:gat} summarizes the GAT-based causal learning procedure that updates node representations and predicts directional causal links.

\begin{algorithm}[!htbp]
\caption{Causal Learning and Link Classification}
\label{alg:gat}
\begin{algorithmic}[1]
\Require Graphs $\{\mathcal{G}\}$, number of GAT layers $L$, number of mini-batches $B$, threshold $\delta$
\Ensure Predicted causal links $\mathcal{C}$

\For{each mini-batch $b = 1$ to $B$}
    \For{each GAT layer $l = 1$ to $L$}
        \State Update node representations using attention-based message passing.
    \EndFor
    \State Compute causal scores for candidate event pairs.
    \State Update model parameters using Focal Loss.
\EndFor

\State Identify causal links using threshold $\delta$.
\State \Return predicted causal links $\mathcal{C}$
\end{algorithmic}
\end{algorithm}

We employ a two-layer Graph Attention Network (GAT) to model event interactions and capture causal dependencies across spatial, temporal, and contextual relations. 
Given node features $\mathbf{x}_v$, the GAT updates them through attention-weighted message passing as follows:
\begin{equation}
    \mathbf{h}_v^{(l+1)} = \sigma \left( \sum_{u \in \mathcal{N}(v)} \alpha_{vu}^{(l)} \mathbf{W}^{(l)} \mathbf{h}_u^{(l)} \right),
\end{equation}
where $\mathcal{N}(v)$ denotes the set of neighbors of node $v$, 
$\mathbf{h}_u^{(l)}$ is the node embedding at layer $l$ (with $\mathbf{h}_u^{(0)} = \mathbf{x}_u$), 
$\mathbf{W}^{(l)}$ is a trainable transformation matrix, and 
$\alpha_{vu}^{(l)}$ is the learned attention coefficient. 
The nonlinearity $\sigma(\cdot)$ is the ELU activation function.

Attention coefficients are computed as:
\begin{equation}
    \alpha_{vu}^{(l)} = \text{softmax}_u\!\left(
        \text{LeakyReLU}\!\big(\mathbf{a}^\top\,\mathbf{z}_{vu}^{(l)}\big)
    \right),
\end{equation}
where $\mathbf{z}_{vu}^{(l)} = [\mathbf{W}^{(l)}\mathbf{h}_v^{(l)} \Vert \mathbf{W}^{(l)}\mathbf{h}_u^{(l)}]$ and 
$\mathbf{a}$ is a trainable attention vector. 
Residual connections are applied between layers to stabilize training and maintain hierarchical consistency, as shown in Figure~\ref{fig:framework}.

After the GAT layers, the final node embeddings of event pairs $(v_i, v_j)$ are concatenated and passed into a fully connected classifier to estimate causal likelihood:
\begin{equation}
    \hat{y}_{ij} = \text{softmax}\!\left( f_\text{MLP}([\mathbf{h}_{v_i} \Vert \mathbf{h}_{v_j}]) \right),
\end{equation}
where $f_\text{MLP}(\cdot)$ is a multi-layer perceptron that outputs class logits. 
The resulting distribution $\hat{y}_{ij}$ indicates the predicted probability of a causal or non-causal relation between the event pair.

The model is trained using the Focal Loss~\cite{focalloss}, which mitigates class imbalance by down-weighting easy negatives and emphasizing harder, misclassified causal links:
\begin{equation}
    L_{\text{focal}}(p_t) = - \alpha_t (1 - p_t)^\gamma \log(p_t),
\end{equation}
where $p_t$ is the predicted probability corresponding to the ground-truth class, 
$\alpha_t$ is a balancing factor, and 
$\gamma$ is the focusing parameter.

By jointly training with focal loss, CaST becomes more sensitive to minority causal cases and effectively reduces the impact of abundant non-causal pairs. 
This GAT-based causal learner corresponds to the ``Training Multi-Head GAT Model'' stage in Figure~\ref{fig:framework}, integrating event-level semantics with spatial and temporal context for robust causal detection in disaster-related tweets.

\subsection{Causal Link Classification}
The final stage of CaST identifies directed causal relations between event pairs represented in the constructed graph. 
Each candidate pair is encoded by concatenating the learned event embeddings, which capture both contextual information and their structural roles within the graph. 
A fully connected classifier predicts whether a causal link exists between the two events, producing a directed causal graph that reflects underlying disaster propagation patterns across space and time.

\section{Experiments}
\subsection{Dataset}
We conduct experiments on an in-house dataset consisting of a total of $\sim$167K disaster-related tweets, each annotated with causal relations between events. The raw corpus originally contained several million tweets collected between August~26 and August~30,~2017, corresponding to the Hurricane Harvey disaster period. To construct a manageable and chronologically consistent dataset, we applied a sampling rate of 1:50, preserving temporal order. Relevant attributes such as tweet text, timestamp, and location information were filtered and stored in JSON format following the schema shown in Figure \ref{fig:json_example}.
\begin{figure}[!htbp]
\centering
\begin{lstlisting}[language=json, basicstyle=\ttfamily\small, breaklines=true, frame=single]
{
  "tweet_text": "<tweet text here>",
  "tokens": ["<token_1>", "<token_2>", "..."],
  "events": [
    {"id": "evt_XXX", "trigger": "<trigger_word>", "arguments": {"arg_1": "<argument_1>"}},
    {"id": "evt_YYY", "trigger": "<trigger_word>", "arguments": {"arg_1": "<argument_1>"}}
  ],
  "causal_relation": {
    "relation": <true_or_false>,
    "pairs": [{"CAUSE": "evt_XXX", "EFFECT": "evt_YYY"}]
  },
  "mask": ["O", "I-C", "I-E", "..."],
  "tweet_id": "<tweet_id>",
  "date_str": "<date_string>",
  "date_numeric": "<unix_timestamp>",
  "geolocation": "<location_string>",
  "bounding_box": "(<lat1>,<lon1>),(<lat2>,<lon2>),(<lat3>,<lon3>),(<lat4>,<lon4>)"
}
\end{lstlisting}
\caption{JSON schema of the annotated disaster tweet dataset used in CaST. 
Each tweet includes the identification number, detected events, causal relations, spatial and temporal metadata, 
and token-level causal role masks.}
\label{fig:json_example}
\end{figure}
The dataset was annotated following the MAVEN-ERE event schema \cite{mavenere}, with minor adjustments for our task. Named Entity Recognition (NER) and linguistic pattern matching were first applied to extract candidate event triggers from tweets, which were then manually verified by two expert annotators. Annotators further examined each tweet to label directed causal links between events, specifying the cause–effect pairs. At the token level, a mask field marks causal roles using I-C (inside a cause span), I-E (inside an effect span), and O (outside any causal span). The final dataset includes: (1) event triggers and arguments, (2) temporal and geospatial metadata, (3) intra-tweet causal relations, and (4) span-level causal masks. This structure enables both event-level and edge-level causal discovery under the proposed CaST framework.

Although the CaST framework is capable of modeling inter-tweet causal dependencies through its spatio-temporal graph design, our in-house dataset currently annotates only intra-tweet causal relations. Consequently, this study focuses on evaluating intra-tweet causal discovery performance, while cross-tweet (inter-tweet) causal reasoning is left for future work.

\subsection{Experimental Setup}
We trained the CaST framework on the disaster tweet dataset using an 80--10--10 split for training, validation, and testing, respectively. Each event graph was treated as a sample, and we employed mini-batch training with a batch size of 32 and a learning rate of 0.001. The model was trained for up to 100 epochs on an NVIDIA RTX~5000 Ada Generation GPU with 16\,GB of memory. To address class imbalance between causal and non-causal pairs, we computed class weights based on their frequency ratio in the training data. The input feature dimension corresponds to the concatenated spatio-temporal embedding of each event node. The GNN backbone uses a hidden dimension of 256, 16 attention heads, and a dropout rate of 0.1. Model parameters were optimized using Adam, and training was stopped early if the validation loss did not improve for 10 consecutive epochs. The model achieving the lowest validation loss was selected for final evaluation on the held-out test set. 

\subsection{Baselines}
We compare CaST against a range of representative baselines spanning classical machine learning, neural architectures, transformer models, and graph-based causal discovery methods. To ensure a fair comparison, all baselines operate on the same event–pair representation. For each pair of events $(e_i, e_j)$, we extract their trigger words and construct an input sequence of the form:
\[
\texttt{trigger}_i \; \texttt{[SEP]} \; \texttt{trigger}_j \; \texttt{[SEP]} \; \texttt{tweet\_text}.
\]
The CrisisTransformer embeddings of each event trigger are used whenever semantic embeddings are required. Spatial and temporal features are excluded for the baseline models to isolate the contribution of spatio–temporal reasoning in CaST.

\begin{itemize}

    \item \textbf{Additive Noise Model (ANM)} \cite{anm}.  
    A classical statistical causal discovery method that assumes the effect is a nonlinear function of the cause plus independent noise.

    \item \textbf{Random Forest (RF).}  
    We train a Random Forest classifier using the concatenated CrisisTransformer embeddings of $(e_i, e_j)$ as features. The model uses 100 trees, a maximum depth of~10, \texttt{class\_weight=balanced}, and Gini impurity. These settings follow standard practice and were validated on the development set.

    \item \textbf{Support Vector Machine (SVM).}  
    The SVM baseline uses an RBF kernel with $C=1.0$ and $\gamma=0.01$. The classifier takes the same concatenated event embeddings as input. The model is trained on the same 70--15--15 split as CaST.

    \item \textbf{BiLSTM + Attention} \cite{bilstmatt}.  
    A bidirectional LSTM with two layers and hidden size 256 is used to encode the token sequence. An attention layer aggregates hidden states into an event–pair representation, followed by a 2-layer MLP classifier. We train the model for up to 15~epochs with learning rate $10^{-3}$ and Adam optimizer, selecting the best checkpoint based on validation F1.

    \item \textbf{BERT} \cite{bert}.  
    We fine-tune \texttt{bert-base-uncased} for causal classification. The input sequence is tokenized as described above, and the model pools three signals: the \texttt{[CLS]} embedding, the mean embedding of event$_i$, and the mean embedding of event$_j$. These three vectors are concatenated and passed to a feed-forward classifier. BERT is trained with AdamW, learning rate $2\times 10^{-5}$, batch size 16, and early stopping based on validation F1.

    \item \textbf{Document-level Event Causality Identification (DECI)} \cite{gcn}.  
    A graph-based neural architecture that applies graph convolutional layers over document-level event graphs. We follow the hyperparameter settings reported in the original paper: two GCN layers with hidden size 256 and learning rate $10^{-3}$.

    \item \textbf{Progressive Graph Pairwise Attention Network (PPAT)} \cite{ppat}.  
    A neural architecture that refines event–pair representations through progressive attention layers. We use the authors’ implementation with default hyperparameters, including hierarchical attention and pairwise refinement across layers.

    \item \textbf{DAPrompt} \cite{daprompt}.  
    A prompt-based approach that fine-tunes a pretrained language model using deterministic assumption prompts. We use the recommended configuration from the original paper with learning rate $3\times 10^{-5}$ and batch size 16.

\end{itemize}

For all neural baselines, training is performed with early stopping on validation F1 rather than a fixed epoch budget, which typically results in fewer than 30~epochs. The best model checkpoint is used for evaluation.

\subsection{Evaluation Metrics}
Framework performance was evaluated using five standard classification metrics: Accuracy, Precision, Recall, F1-score, and the Area Under the ROC Curve (AUC). Accuracy measures the overall correctness of predictions, while Precision and Recall capture the model’s ability to identify true causal links without introducing false positives or missing actual relations. The F1-score, as their harmonic mean, provides a balanced indicator of model performance under class imbalance. AUC quantifies the trade-off between true positive and false positive rates, reflecting the model’s discriminative capability across different thresholds.

\subsection{Results}
\subsubsection{Baseline Comparison}
Table \ref{tab:baseline} summarizes the performance of CaST and existing baselines on the disaster tweet dataset. Overall, CaST achieves the best results across most metrics, attaining 0.87 accuracy, 0.84 precision, 0.85 recall, and 0.85 F1-score. These consistent gains confirm the effectiveness of incorporating spatial and temporal signals into event representations for causal discovery.

\begin{table}[!htbp]
\centering
\caption{Performance comparison on the in-house dataset.}
\label{tab:baseline}
\resizebox{\linewidth}{!}{
\begin{tabular}{l|ccccc}
\toprule
\textbf{Model} & \textbf{Acc.} & \textbf{Prec.} & \textbf{Rec.} & \textbf{F1} & \textbf{AUC} \\
\midrule
ANM \cite{anm} & 0.50 & 0.50 & 0.53 & 0.52 & 0.50 \\
RF & 0.71 & 0.65 & 0.69 & 0.65 & 0.68 \\
SVM & 0.63 & 0.63 & 0.68 & 0.60 & 0.68 \\
BiLSTM+Att. \cite{bilstmatt} & 0.70 & 0.60 & 0.62 & 0.61 & 0.65 \\
BERT \cite{bert} & 0.79 & 0.74 & 0.81 & 0.75 & \textbf{0.89} \\
DECI \cite{gcn} & 0.54 & 0.62 & 0.79 & 0.70 & 0.58 \\
PPAT \cite{ppat} & 0.86 & 0.60 & 0.30 & 0.40 & 0.88 \\
DAPrompt \cite{daprompt} & 0.77 & 0.52 & 0.78 & 0.62 & 0.86 \\
\midrule
\textbf{CaST (Ours)} & \textbf{0.87} & \textbf{0.84} & \textbf{0.85} & \textbf{0.85} & 0.85 \\
\bottomrule
\end{tabular}
}
\end{table}

Among the baselines, transformer-based methods such as BERT and DAPrompt outperform earlier neural and statistical approaches. BERT achieves an F1-score of 0.75 and the highest AUC of 0.89, indicating its strong semantic understanding of contextual relations between event pairs. However, it lacks explicit modeling of event occurrence in space and time, which limits its ability to capture causal propagation patterns typical of disaster scenarios. DAPrompt performs comparably (F1 = 0.62, AUC = 0.86) due to its use of adaptive prompting, but its precision remains low, suggesting frequent false positives when causal cues are implicit or ambiguous.

Graph-based models, including DECI and PPAT, demonstrate mixed performance. DECI achieves high recall (0.79) but low accuracy (0.54), reflecting its tendency to over-predict causal connections. PPAT, conversely, attains the highest baseline accuracy (0.86) but at the cost of poor recall (0.30), indicating that its progressive refinement captures only clear, surface-level causality while missing more subtle links. Traditional machine learning baselines (SVM, RF) perform moderately well, showing that lexical and syntactic cues can offer limited discriminative power, while ANM performs near random due to its reliance on numerical assumptions unsuitable for textual input. Importantly, all of the baselines above hardly achieve a balanced trade-off between precision and recall. Most of them tend to favor one over the other, either missing critical causal links or introducing excessive false positives. This imbalance indicates that existing methods often overlook important causal information dispersed across space and time, leading to incomplete understanding of event propagation in disasters.

In contrast, CaST effectively balances both semantic and structural reasoning. By embedding spatial and temporal features directly into event representations before graph learning, it captures dependencies that span across locations and time intervals—an aspect particularly relevant in disaster contexts where cascading effects often follow a spatial-temporal chain (e.g., flooding leading to power outages). This integration enables CaST to reduce false positives compared to BERT and improve recall compared to PPAT, achieving robust generalization on real-world disaster narratives. Quantitatively, CaST attains both high precision (0.84) and recall (0.85), demonstrating a well-balanced capability to identify true causal links without over-predicting non-causal ones. This balance is particularly crucial for disaster response, where false positives can mislead situational understanding and false negatives may overlook critical event dependencies. These results highlight that causal reasoning in social media text benefits significantly from structured spatio-temporal context rather than relying solely on semantic cues or static graph reasoning. CaST’s improvements demonstrate the value of unifying spatial, temporal, and semantic representations within a single graph-based learning framework for disaster-related causal discovery.

It is worth noting that the dataset is inherently imbalanced, with causal event pairs forming only a small fraction of all possible event combinations (approximately 1:3 in this in-house dataset). As a result, several baseline models exhibit relatively high accuracy but noticeably lower precision, recall, and F1 scores. This occurs because such models are prone to the imbalance in the data, tending to predict the majority non-causal class more frequently. Consequently, accuracy alone can be misleading in this setting, while F1 and AUC serve as more reliable indicators of causal detection performance, as they better reflect the model’s ability to identify true causal links amid class imbalance. To further examine the contribution of each component in CaST, we conduct an ablation study that isolates the effects of spatial and temporal features, as discussed in the next subsection.

\subsection{Ablation Study}
\begin{figure}[!htbp]
    \centering
    \includegraphics[width=\linewidth]{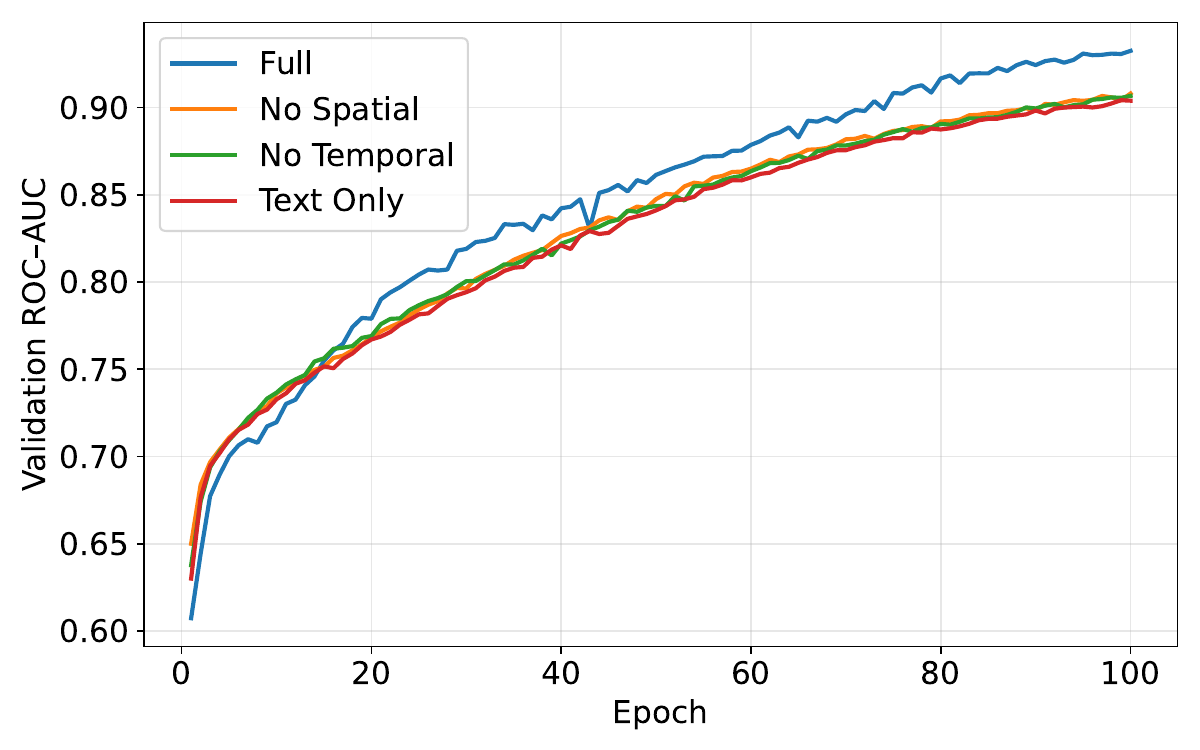}
    \caption{ROC–AUC comparison of CaST variants during training.}
    \vspace{-2mm}
    \label{fig:ablation_roc}
\end{figure}

\begin{figure}[!htbp]
    \centering
    \includegraphics[width=\linewidth]{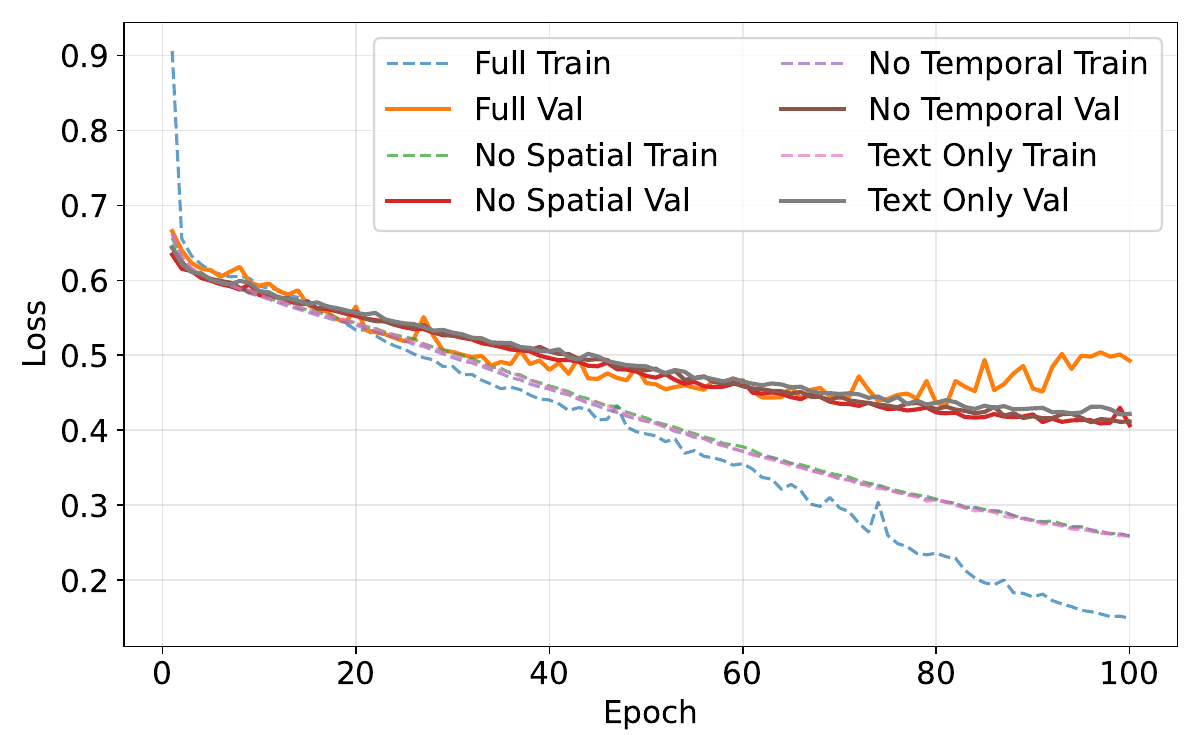}
    \caption{Training and validation loss curves for CaST and its ablated variants.}
    \label{fig:ablation_loss}
\end{figure}

To evaluate the contribution of spatial and temporal reasoning in CaST, we perform a series of ablation experiments and compare four configurations: without both spatial and temporal components (text-only), without spatial, without temporal, and the full model. The quantitative results are summarized in Table \ref{tab:ablation}, while the corresponding learning trends are shown in Figure \ref{fig:ablation_roc} and Figure \ref{fig:ablation_loss}.
\begin{table}[!htbp]
\centering
\caption{Comparison of CaST variants on the in-house dataset.}
\label{tab:ablation}
\begin{tabular}{l|ccccc}
\toprule
\textbf{Configuration} & \textbf{Acc} & \textbf{Prec.} & \textbf{Recall} & \textbf{F1} & \textbf{AUC} \\
\midrule
w/o Spatial \& Temporal & 0.81 & 0.79 & 0.78 & 0.79 & 0.81 \\
w/o Temporal            & 0.83 & 0.80 & 0.79 & 0.80 & 0.83 \\
w/o Spatial             & 0.84 & 0.80 & 0.78 & 0.79 & 0.82 \\
\textbf{Full (CaST)}    & \textbf{0.87} & \textbf{0.84} & \textbf{0.85} & \textbf{0.85} & \textbf{0.85} \\
\bottomrule
\end{tabular}
\end{table}

\textbf{Ablation results.} The base variant without either component (\textit{w/o Spatial \& Temporal}) shows the lowest performance across all metrics (F1 = 0.79), underscoring that semantic embeddings alone are insufficient for representing the dynamic and context-dependent nature of disaster events. Introducing spatial or temporal information individually yields measurable improvements of about 1–2\% across most metrics, indicating that each dimension contributes complementary information to causal reasoning. The model without temporal information (\textit{w/o Temporal}) performs slightly worse than the one without spatial cues (\textit{w/o Spatial}), which aligns with the principle that causal relationships often follow a chronological sequence—events that occur earlier are more likely to cause later ones. Temporal modeling thus enhances recall by identifying ordered event dependencies, while spatial modeling strengthens precision by filtering unrelated events that may co-occur temporally but occur in distinct geographic regions.

When both spatial and temporal dimensions are integrated, the full CaST model achieves the highest overall performance (Acc = 0.87, F1 = 0.85, AUC = 0.85), showing a balanced trade-off between precision and recall. This balanced improvement again highlights the synergy between spatial proximity and temporal progression: together they enable the model to capture realistic causal propagation patterns that unfold over space and time during disasters. These findings affirm that causal reasoning in the disaster domain cannot rely solely on semantic cues or pairwise correlations but must jointly encode spatio-temporal dependencies to achieve robust and interpretable causal discovery.

\textbf{Training Stability and Generalization Trends.} Figure \ref{fig:ablation_roc} compares the validation ROC–AUC of all configurations across training epochs. The full CaST model consistently outperforms its ablated counterparts, with the performance gap widening steadily after around 40 epochs. This trend indicates that spatial and temporal cues provide complementary context that strengthens causal discrimination as training progresses. In contrast, models lacking these features tend to plateau earlier, suggesting limited capacity to capture long-range or cross-event dependencies.
To further understand this behavior, Figure \ref{fig:ablation_loss} examines the convergence dynamics across configurations. The full CaST variant not only achieves the lowest training loss but also converges more smoothly, while the ablated models exhibit minor oscillations, indicating overfitting and unstable optimization. This stability implies that spatio-temporal information acts as a structural regularizer, guiding the model to learn more coherent event dependencies across time and space.

\section{Conclusion}
This paper presented CaST, a novel framework for causal discovery via spatio-temporal graphs in disaster-related tweets. Motivated by the need to understand how disaster effects propagate across regions and time, CaST unifies linguistic, spatial, and temporal cues into a single event-centric graph structure. The framework encodes contextual, temporal, and location-aware dependencies between events extracted from social media messages and learns to infer causal relations through a multi-head Graph Attention Network (GAT). Unlike prior text-only or static graph approaches, CaST explicitly models the dynamic and multi-dimensional nature of disaster events. Each event representation integrates semantic embeddings with temporal ordering and geolocation features, allowing the network to reason over evolving causal chains. The incorporation of focal loss further enhances the model’s robustness under severe class imbalance, emphasizing hard-to-detect causal links that are often overlooked in large-scale social datasets. Extensive experiments on our in-house dataset of 167K disaster tweets show that CaST achieves consistent improvements over state-of-the-art baselines, with superior F1 and AUC scores. In future work, we aim to explore how external or domain-specific knowledge can be leveraged to enrich event representations and improve causal inference. 

\section*{Acknowledgment} This work was supported by NSF - USA CNS-2219614.


\begin{thebibliography}{10}
\providecommand{\url}[1]{#1}
\csname url@samestyle\endcsname
\providecommand{\newblock}{\relax}
\providecommand{\bibinfo}[2]{#2}
\providecommand{\BIBentrySTDinterwordspacing}{\spaceskip=0pt\relax}
\providecommand{\BIBentryALTinterwordstretchfactor}{4}
\providecommand{\BIBentryALTinterwordspacing}{\spaceskip=\fontdimen2\font plus
\BIBentryALTinterwordstretchfactor\fontdimen3\font minus \fontdimen4\font\relax}
\providecommand{\BIBforeignlanguage}[2]{{%
\expandafter\ifx\csname l@#1\endcsname\relax
\typeout{** WARNING: IEEEtran.bst: No hyphenation pattern has been}%
\typeout{** loaded for the language `#1'. Using the pattern for}%
\typeout{** the default language instead.}%
\else
\language=\csname l@#1\endcsname
\fi
#2}}
\providecommand{\BIBdecl}{\relax}
\BIBdecl

\bibitem{gat}
\BIBentryALTinterwordspacing
P.~Veli{\v{c}}kovi{\'{c}}, G.~Cucurull, A.~Casanova, A.~Romero, P.~Li{\`{o}}, and Y.~Bengio, ``{Graph Attention Networks},'' \emph{International Conference on Learning Representations}, 2018, accepted as poster. [Online]. Available: \url{https://openreview.net/forum?id=rJXMpikCZ}
\BIBentrySTDinterwordspacing

\bibitem{khoo1998}
C.~S.~G. Khoo, J.~Kornfilt, R.~N. Oddy, and S.~H. Myaeng, ``Automatic extraction of cause-effect information from newspaper text without knowledge-based inferencing,'' \emph{Literary and Linguistic Computing}, vol.~13, no.~4, pp. 177--186, 1998.

\bibitem{blanco2008}
E.~Blanco, N.~Castell, and D.~Moldovan, ``Causal relation extraction,'' in \emph{Proc. of the 6th International Conference on Language Resources and Evaluation (LREC’08)}, 2008.

\bibitem{girju2003}
R.~Girju, ``Automatic detection of causal relations for question answering,'' in \emph{Proc. of the ACL 2003 Workshop on Multilingual Summarization and Question Answering}, 2003, pp. 76--83.

\bibitem{cao2021}
P.~Cao, X.~Zuo, Y.~Chen, K.~Liu, J.~Zhao, Y.~Chen, and W.~Peng, ``Knowledge-enriched event causality identification via latent structure induction networks,'' in \emph{Proc. of the 59th Annual Meeting of the Association for Computational Linguistics (ACL 2021)}, 2021, pp. 4476--4488.

\bibitem{shen2022}
S.~Shen, H.~Zhou, T.~Wu, and G.~Qi, ``Event causality identification via derivative prompt joint learning,'' in \emph{Proc. of the 29th International Conference on Computational Linguistics (COLING 2022)}, 2022, pp. 2288--2299.

\bibitem{zuo2021}
X.~Zuo, P.~Cao, Y.~Chen, K.~Liu, J.~Zhao, W.~Peng, and Y.~Chen, ``{L}earn{DA}: Learnable knowledge-guided data augmentation for event causality identification,'' in \emph{Proceedings of the 59th Annual Meeting of the Association for Computational Linguistics and the 11th International Joint Conference on Natural Language Processing (Volume 1: Long Papers)}, C.~Zong, F.~Xia, W.~Li, and R.~Navigli, Eds.\hskip 1em plus 0.5em minus 0.4em\relax Online: Association for Computational Linguistics, Aug. 2021, pp. 3558--3571.

\bibitem{anuyah2025}
\BIBentryALTinterwordspacing
S.~Anuyah, J.~Vanschaik, P.~Jain, S.~Lehman, and S.~Chakraborty, ``An empirical study of causal relation extraction transfer: Design and data,'' 2025. [Online]. Available: \url{https://arxiv.org/abs/2503.06076}
\BIBentrySTDinterwordspacing

\bibitem{chun2025}
Y.~Chun, S.~Ha, J.~Bai \emph{et~al.}, ``Utilizing pre-trained language models for data augmentation task of event causality identification,'' \emph{Research Square}, May 2025, preprint (Version 1).

\bibitem{gao2019}
L.~Gao, P.~K. Choubey, and R.~Huang, ``Modeling document-level causal structures for event causal relation identification,'' in \emph{Proc. of NAACL-HLT 2019 (Volume 1: Long Papers)}, 2019, pp. 1808--1817.

\bibitem{gcn}
M.~Tran~Phu and T.~H. Nguyen, ``Graph convolutional networks for event causality identification with rich document-level structures,'' in \emph{Proceedings of the 2021 Conference of the North American Chapter of the Association for Computational Linguistics: Human Language Technologies}, K.~Toutanova, A.~Rumshisky, L.~Zettlemoyer, D.~Hakkani-Tur, I.~Beltagy, S.~Bethard, R.~Cotterell, T.~Chakraborty, and Y.~Zhou, Eds.\hskip 1em plus 0.5em minus 0.4em\relax Online: Association for Computational Linguistics, Jun. 2021, pp. 3480--3490.

\bibitem{fan2022}
K.~Fan, P.~Li, Y.~Du, D.~Yu, Z.~Yu, Y.~Zhang, and W.~Lu, ``Document-level event causality identification via heterogeneous graph attention networks,'' in \emph{Proc. of the 60th Annual Meeting of the Association for Computational Linguistics (ACL 2022)}, 2022, pp. 1252--1263.

\bibitem{identifying}
\BIBentryALTinterwordspacing
C.~Liu, W.~Xiang, and B.~Wang, ``Identifying while learning for document event causality identification,'' in \emph{Proceedings of the 62nd Annual Meeting of the Association for Computational Linguistics (Volume 1: Long Papers)}, L.-W. Ku, A.~Martins, and V.~Srikumar, Eds.\hskip 1em plus 0.5em minus 0.4em\relax Bangkok, Thailand: Association for Computational Linguistics, Aug. 2024, pp. 3815--3827. [Online]. Available: \url{https://aclanthology.org/2024.acl-long.210/}
\BIBentrySTDinterwordspacing

\bibitem{pu2024}
R.~Pu, Y.~Li, J.~Zhao, S.~Wang \emph{et~al.}, ``A joint framework with heterogeneous-relation-aware graph and multi-channel label enhancing strategy for event causality extraction,'' in \emph{Proc. of the 38th AAAI Conference on Artificial Intelligence (AAAI 2024)}, 2024, pp. 18\,879--18\,887.

\bibitem{hu2023}
Z.~Hu, Z.~Li, X.~Jin, L.~Bai, S.~Guan, J.~Guo, and X.~Cheng, ``Semantic structure enhanced event causality identification,'' in \emph{Proc. of the 61st Annual Meeting of the Association for Computational Linguistics (ACL 2023)}, 2023, pp. 10\,901--10\,913.

\bibitem{mirza2014}
P.~Mirza and S.~Tonelli, ``An analysis of causality between events and its relation to temporal information,'' in \emph{Proc. of COLING 2014 (25th International Conference on Computational Linguistics)}, 2014, pp. 2097--2106.

\bibitem{ppat}
Z.~Liu, B.~Hu, Z.~Xu, and M.~Zhang, ``Ppat: progressive graph pairwise attention network for event causality identification,'' in \emph{Proceedings of the Thirty-Second International Joint Conference on Artificial Intelligence}, ser. IJCAI '23, 2023.

\bibitem{liu2025}
Z.~Liu, Y.~Liang, and W.~Ni, ``Document-level causal event extraction enhanced by temporal relations using dual-channel neural network,'' \emph{Electronics}, vol.~14, no.~5, p. 992, 2025.

\bibitem{dong2025}
X.~Dong, E.~Mas, B.~Adriano, and S.~Koshimura, ``Towards real-time extraction of cascading effect and spatiotemporal analysis using social media data,'' \emph{International Journal of Disaster Risk Reduction}, vol. 125, p. 105512, 2025.

\bibitem{lenti2025}
J.~Lenti \emph{et~al.}, ``Causal modeling of climate activism on reddit,'' in \emph{Proc. of The Web Conference (WWW) 2025}, 2025, pp. 590--600.

\bibitem{kiciman2024}
E.~Kiciman, R.~Ness, A.~Sharma, and C.~Tan, ``Causal reasoning and large language models: Opening a new frontier for causality,'' \emph{Transactions on Machine Learning Research (TMLR)}, 2024, to appear; arXiv:2305.00050.

\bibitem{long2023}
S.~Long, A.~Piché, V.~Zantedeschi, T.~Schuster, and A.~Drouin, ``Causal discovery with language models as imperfect experts,'' 2023.

\bibitem{shyalika2024}
C.~Shyalika and *, ``Causal event graph-guided language-based spatiotemporal question answering,'' in \emph{Proc. of the AAAI Spring Symposium on Knowledge Graphs for XAI}, 2024.

\bibitem{Holland01121986}
P.~W. Holland, ``Statistics and causal inference,'' \emph{Journal of the American Statistical Association}, vol.~81, no. 396, pp. 945--960, 1986.

\bibitem{crisis}
\BIBentryALTinterwordspacing
R.~Lamsal, M.~R. Read, and S.~Karunasekera, ``Crisistransformers: Pre-trained language models and sentence encoders for crisis-related social media texts,'' \emph{Knowledge-Based Systems}, vol. 296, p. 111916, 2024. [Online]. Available: \url{https://www.sciencedirect.com/science/article/pii/S0950705124005501}
\BIBentrySTDinterwordspacing

\bibitem{focalloss}
\BIBentryALTinterwordspacing
T.~Lin, P.~Goyal, R.~B. Girshick, K.~He, and P.~Doll{\'{a}}r, ``Focal loss for dense object detection,'' \emph{CoRR}, vol. abs/1708.02002, 2017. [Online]. Available: \url{http://arxiv.org/abs/1708.02002}
\BIBentrySTDinterwordspacing

\bibitem{mavenere}
X.~Wang, Y.~Chen, N.~Ding, H.~Peng, Z.~Wang, Y.~Lin, X.~Han, L.~Hou, J.~Li, Z.~Liu, P.~Li, and J.~Zhou, ``Maven-ere: A unified large-scale dataset for event coreference, temporal, causal, and subevent relation extraction,'' in \emph{Proceedings of EMNLP}, 2022.

\bibitem{anm}
P.~Hoyer, D.~Janzing, J.~M. Mooij, J.~Peters, and B.~Sch\"{o}lkopf, ``Nonlinear causal discovery with additive noise models,'' in \emph{Advances in Neural Information Processing Systems}, D.~Koller, D.~Schuurmans, Y.~Bengio, and L.~Bottou, Eds., vol.~21.\hskip 1em plus 0.5em minus 0.4em\relax Curran Associates, Inc., 2008.

\bibitem{bilstmatt}
Z.~Li, Q.~Li, X.~Zou, and J.~Ren, ``Causality extraction based on self-attentive bilstm-crf with transferred embeddings,'' \emph{CoRR}, vol. abs/1904.07629, 2019.

\bibitem{bert}
J.~Devlin, M.-W. Chang, K.~Lee, and K.~Toutanova, ``{BERT}: Pre-training of deep bidirectional transformers for language understanding,'' in \emph{Proceedings of the 2019 Conference of the North {A}merican Chapter of the Association for Computational Linguistics: Human Language Technologies, Volume 1 (Long and Short Papers)}, J.~Burstein, C.~Doran, and T.~Solorio, Eds.\hskip 1em plus 0.5em minus 0.4em\relax Minneapolis, Minnesota: Association for Computational Linguistics, Jun. 2019, pp. 4171--4186.

\bibitem{daprompt}
W.~Xiang, C.~Zhan, Q.~Zhang \emph{et~al.}, ``Daprompt: deterministic assumption prompt learning for event causality identification,'' \emph{Neural Computing and Applications}, vol.~37, pp. 21\,743--21\,759, 2025.

\end{thebibliography}

\end{document}